\documentclass[twocolumn]{aastex631}
\usepackage[varg]{txfonts}
\usepackage{graphicx} 
\usepackage{url} 

\begin{document}
\title{Early Planet Formation in Embedded Disks (eDisk) XV:\\
Influence of Magnetic Field Morphology in Dense Cores on Sizes of Protostellar Disks}

\author[0000-0003-1412-893X]{Hsi-Wei Yen}
\affiliation{Academia Sinica Institute of Astronomy \& Astrophysics,
11F of Astronomy-Mathematics Building, AS/NTU, No.1, Sec. 4, Roosevelt Rd, Taipei 10617, Taiwan}

\author[0000-0001-5058-695X]{Jonathan P. Williams}
\affiliation{Institute for Astronomy, University of Hawai‘i at Mānoa, 2680 Woodlawn Dr., Honolulu, HI 96822, USA}

\author[0000-0003-4361-5577]{Jinshi Sai (Insa Choi)}
\affiliation{Academia Sinica Institute of Astronomy \& Astrophysics,
11F of Astronomy-Mathematics Building, AS/NTU, No.1, Sec. 4, Roosevelt Rd, Taipei 10617, Taiwan}

\author[0000-0003-2777-5861]{Patrick M. Koch}
\affiliation{Academia Sinica Institute of Astronomy \& Astrophysics,
11F of Astronomy-Mathematics Building, AS/NTU, No.1, Sec. 4, Roosevelt Rd, Taipei 10617, Taiwan}

\author[0000-0002-9143-1433]{Ilseung Han}
\affiliation{Division of Astronomy and Space Science, University of Science and Technology, 217 Gajeong-ro, Yuseong-gu, Daejeon 34113, Republic of Korea}
\affiliation{Korea Astronomy and Space Science Institute, 776 Daedeok-daero, Yuseong-gu, Daejeon 34055, Republic of Korea}

\author[0000-0001-9133-8047]{Jes K. J{\o}rgensen}
\affil{Niels Bohr Institute, University of Copenhagen, {\O}ster Voldgade 5--7, DK~1350 Copenhagen K., Denmark}

\author[0000-0003-4022-4132]{Woojin Kwon}
\affiliation{Department of Earth Science Education, Seoul National University, 1 Gwanak-ro, Gwanak-gu, Seoul 08826, Republic of Korea}
\affiliation{SNU Astronomy Research Center, Seoul National University, 1 Gwanak-ro, Gwanak-gu, Seoul 08826, Republic of Korea}

\author[0000-0002-3179-6334]{Chang Won Lee}
\affiliation{Division of Astronomy and Space Science, University of Science and Technology, 217 Gajeong-ro, Yuseong-gu, Daejeon 34113, Republic of Korea}
\affiliation{Korea Astronomy and Space Science Institute, 776 Daedeok-daero, Yuseong-gu, Daejeon 34055, Republic of Korea}

\author[0000-0002-7402-6487]{Zhi-Yun Li}
\affiliation{University of Virginia, 530 McCormick Rd., Charlottesville, Virginia 22904, USA}

\author[0000-0002-4540-6587]{Leslie W.  Looney}
\affiliation{Department of Astronomy, University of Illinois, 1002 West Green St, Urbana, IL 61801, USA}

\author[0000-0002-0554-1151]{Mayank Narang}
\affiliation{Academia Sinica Institute of Astronomy \& Astrophysics,
11F of Astronomy-Mathematics Building, AS/NTU, No.1, Sec. 4, Roosevelt Rd, Taipei 10617, Taiwan}

\author[0000-0003-0998-5064]{Nagayoshi Ohashi}
\affiliation{Academia Sinica Institute of Astronomy \& Astrophysics,
11F of Astronomy-Mathematics Building, AS/NTU, No.1, Sec. 4, Roosevelt Rd, Taipei 10617, Taiwan}

\author[0000-0003-0845-128X]{Shigehisa Takakuwa}
\affiliation{Department of Physics and Astronomy, Graduate School of Science and Engineering, Kagoshima University, 1-21-35 Korimoto, Kagoshima, Kagoshima 890-0065, Japan}
\affiliation{Academia Sinica Institute of Astronomy \& Astrophysics,
11F of Astronomy-Mathematics Building, AS/NTU, No.1, Sec. 4, Roosevelt Rd, Taipei 10617, Taiwan}

\author[0000-0002-6195-0152]{John J. Tobin}
\affil{National Radio Astronomy Observatory, 520 Edgemont Rd., Charlottesville, VA 22903 USA} 

\author[0000-0003-4518-407X]{Itziar de Gregorio-Monsalvo}
\affiliation{European Southern Observatory, Alonso de Cordova 3107, Casilla 19, Vitacura, Santiago, Chile}

\author[0000-0001-5522-486X]{Shih-Ping Lai}
\affiliation{Institute of Astronomy, National Tsing Hua University, No. 101, Section 2, Kuang-Fu Road, Hsinchu 30013, Taiwan}
\affiliation{Center for Informatics and Computation in Astronomy, National Tsing Hua University, No. 101, Section 2, Kuang-Fu Road, Hsinchu 30013, Taiwan}
\affiliation{Department of Physics, National Tsing Hua University, No. 101, Section 2, Kuang-Fu Road, Hsinchu 30013, Taiwan}
\affiliation{Academia Sinica Institute of Astronomy \& Astrophysics,
11F of Astronomy-Mathematics Building, AS/NTU, No.1, Sec. 4, Roosevelt Rd, Taipei 10617, Taiwan}

\author[0000-0003-3119-2087]{Jeong-Eun Lee}
\affiliation{Department of Physics and Astronomy, Seoul National University, 1 Gwanak-ro, Gwanak-gu, Seoul 08826, Korea}

\author[0000-0001-8105-8113]{Kengo Tomida}
\affiliation{Astronomical Institute, Graduate School of Science, Tohoku University, Sendai 980-8578, Japan}

\correspondingauthor{Hsi-Wei Yen}
\email{hwyen@asiaa.sinica.edu.tw}

\shorttitle{eDisk XV. Disk Sizes and Magnetic Field Morphology in Dense Cores}
\shortauthors{Yen et al.}

\begin{abstract}
The magnetic field of a molecular cloud core may play a role in the formation of circumstellar disks in the core.
We present magnetic field morphologies in protostellar cores of 16 targets in the Atacama Large Millimeter/submillimeter Array large program ``Early Planet Formation in Embedded Disks (eDisk)'', which resolved their disks with 7~au resolutions. 
The 0.1-pc scale magnetic field morphologies were inferred from the James Clerk Maxwell Telescope (JCMT) POL-2 observations.  
The mean orientations and angular dispersions of the magnetic fields in the dense cores are measured and compared with the radii of the 1.3\,mm continuum disks and the dynamically determined protostellar masses from the eDisk program.
We observe a significant correlation between the disk radii and the stellar masses. 
We do not find any statistically significant dependence of the disk radii on the projected misalignment angles between the rotational axes of the disks and the magnetic fields in the dense cores, nor on the angular dispersions of the magnetic fields within these cores. 
However, when considering the projection effect, we cannot rule out a positive correlation between disk radii and misalignment angles in three-dimensional space. 
Our results suggest that the morphologies of magnetic fields in dense cores do not play a dominant role in the disk formation process. 
Instead, the sizes of protostellar disks may be more strongly affected by the amount of mass that has been accreted onto star+disk systems, and possibly other parameters, for example, magnetic field strength, core rotation, and magnetic diffusivity.
\end{abstract}

\keywords{Star formation (1569), Interstellar magnetic fields (845), Star forming regions (1565), Protostars (1302), Circumstellar disks (235)}

\section{Introduction}
Protostellar disks around Class 0 and I protostars can be more massive than protoplanetary disks around pre-main sequence stars \citep{Tychoniec20,Sheehan22}, 
making them potential sites for planet formation \citep{Manara18}. 
Based on hydrodynamics (HD), protostellar disks form and grow rapidly due to the conservation of angular momentum during the collapse of dense cores \citep{Terebey84}. 
However, when considering magnetohydrodynamics (MHD), magnetic fields can suppress disk formation and growth \citep{Li14, Tsukamoto23}. 
The efficiency of magnetic braking depends on turbulence and magnetic field orientation, diffusivity, and strength in dense cores \citep{Hirano20,Lee21}. 
While these physical parameters have been theoretically studied, their impacts on disk formation and evolution have not yet been conclusively quantified observationally with the previous attempts \citep{Galametz20, Yen21a, Gupta22}.

The Atacama Large Millimeter/submillimeter Array (ALMA) large program, ``Early Planet Formation in Embedded Disks (eDisk)'', systematically surveys 19 Class 0 and I protostars at an angular resolution of 0$\farcs$04 ($\sim$7\,au) in the continuum and molecular lines at 1.3\,mm, 
and provides a large uniform sample of resolved protostellar disks to study their disk structures and kinematics \citep{Ohashi23}.
To investigate the influence of the magnetic fields in the dense cores on their protostellar disks, 
we observed submillimeter polarized dust emission in the eDisk targets with the James Clerk Maxwell Telescope (JCMT) using its polarimeter POL-2 and bolometer camera SCUBA-2 \citep{Holland13}.
The magnetic field structures on a 0.1\,pc scale observed with JCMT can be a proxy of the initial configuration of the magnetic fields in the dense cores to compare with MHD simulations. 
In the simulations, the initial orientation and structures of the magnetic fields, which can be subsequently dragged and twisted in the inner dynamically collapsing region, have a profound impact on the formation and evolution of protostellar disks \citep{Hennebelle09,Joos12,Li13,Hirano20}.

Here we present the JCMT POL-2 850\,$\mu$m data of the eDisk targets (Section~\ref{sec:obs}), 
and measure the mean orientations and angular dispersions of the magnetic fields in the dense cores on a 0.1\,pc scale (Section~\ref{sec:results}). 
We then compare the disk radii and stellar masses measured from the eDisk program with the inferred magnetic field structures from the JCMT data (Section~\ref{sec:analysis}). 
By examining the correlations between the disk radii, stellar masses, and orientations and angular dispersions of the magnetic fields, 
we discuss the dependences of the disk radii on these parameters and their relative importance in the formation and evolution of protostellar disks (Section~\ref{sec:discussion}).

\begin{deluxetable*}{lcccccccccc}
\tablecaption{Sample sources and their properties} \label{tab:sample}
\tablewidth{0pt}
\tablehead{
\colhead{Source name} & \colhead{ICRS R.A.} & \colhead{ICRS Dec.} & \colhead{Class} & \colhead{Distance} & \colhead{$T_{\rm bol}$} & 
\colhead{$L_{\rm bol}$} & \colhead{$M_{\star}$} & \colhead{$R_{\rm disk}$} & \colhead{$\theta_{\rm rot}$} & \colhead{Ref.} \\
\colhead{} & \colhead{} & \colhead{} & \colhead{} &
\colhead{(pc)} & \colhead{(K)} & \colhead{($L_\sun$)} & \colhead{($M_\sun$)} & \colhead{(au)} & \colhead{(\arcdeg)} &  \colhead{}
}
\startdata
L1489~IRS & 04:04:43.080 & +26:18:56.12 & I & 146 & 213 & 3.4 & 1.7$\pm$0.2 & 485 & 156 & 1, 2\\
IRAS~04166+2706 & 04:19:42.505 & +27:13:35.83 & 0 & 156 & 61 & 0.40 & 0.27$\pm$0.12 &18 & 32 & 1, 3\\
IRAS~04169+2702 & 04:19:58.477 & +27:09:56.82 & I & 156 & 163 & 1.5 & 1.4$\pm$0.7 & 29 & 49 & 1, 4\\
IRAS~04302+2247 & 04:33:16.499 & +22:53:20.23 & I & 160 & 88 & 0.43 & 1.6$\pm$0.4 & 292 & 85 & 1, 5\\
L1527~IRS & 04:39:53.878 & +26:03:09.43 & 0 & 140 & 41 & 1.3 & 0.4$\pm$0.1 & 53 & 92 & 1, 6\\
Ced110~IRS4A\tablenotemark{a} & 11:06:46.369 & $-$77:22:32.88 & 0 & 189 & 68 & 1.0 & 1.33$\pm$0.12 & 60 & 14 & 1, 7\\
Ced110~IRS4B\tablenotemark{a} & 11:06:46.772 & $-$77:22:32.76 & 0 &  189 & 68 & 10 & 0.04$\pm$0.02 & 22 & 175 & 1, 7\\
BHR71~IRS2\tablenotemark{a} & 12:01:34.008 & $-$65:08:48.08 & 0 & 176 & 39 & 1.1 & 0.25 & 7 & 158 & 1, 8\\
BHR71~IRS1\tablenotemark{a} & 12:01:36.476 & $-$65:08:49.37 & 0 & 176 & 66 & 10 & 0.4 & 42 & 8 & 1, 8\\
IRAS~15398$-$3359 & 15:43:02.232 & $-$34:09:06.96 & 0 & 155 & 50 & 1.4 & 0.06$\pm$0.04 & 4 & 27 & 1, 9\\
GSS30~IRS3 & 16:26:21.715 & $-$24:22:51.09 & 0 & 138 & 50 & 1.7 & 0.46$\pm$0.14 & 64 & 19 & 1, 10\\
Oph~IRS43~VLA1 & 16:27:26.906 & $-$24:40:50.81 & I & 137 & 193 & 4.1 & 1.0$\pm$0.3 & 10 & 46& 1, 11\\
Oph~IRS43~VLA2 & 16:27:26.911 & $-$24:40:51.40 & I & 137 & 193 & 4.1 & 1.0$\pm$0.3 & 2 & 41 & 1, 11\\
IRAS~16253$-$2429 & 16:28:21.615 & $-$24:36:24.33 & 0 & 139 & 42 & 0.16 & 0.15$\pm$0.02 & 13 & 23 & 1, 12\\
Oph~IRS63 & 16:31:35.654 & $-$24:01:30.08 & I & 132 & 348 & 1.3 & 0.5$\pm$0.2 & 48 & 59 & 1, 13\\
IRAS~16544$-$1604 & 16:57:19.643 & $-$16:09:24.02 & 0 & 151 & 50 & 0.89 & 0.14 & 27 & 135 & 1, 14\\
R~CrA~IRS5N & 19:01:48.480 & $-$36:57:15.39 & 0 & 147 & 59 & 1.4 & 0.29$\pm$0.11 & 47 & 171 & 1, 15\\
R~CrA~IRS7B-a & 19:01:56.420 & $-$36:57:28.66 & I & 152 & 88 & 5.1 & 2.65$\pm$0.55 & 60 & 25 & 1\\
R~CrA~IRS7B-b & 19:01:56.385 & $-$36:57:28.11 & I & 152 & 88 & 5.1 & \nodata & 24 & 25 & 1\\
R~CrA~IRAS~32A & 19:02:58.722 & $-$37:07:37.39 & 0 & 150 & 64 & 1.6 & 0.71$\pm$0.2 & 27 & 45 & 1, 16\\
R~CrA~IRAS~32B & 19:02:58.642 & $-$37:07:36.39 & 0 & 150 & 64 & 1.6 & 0.4$\pm$0.08 & 23 & 42 & 1, 16\\
TMC-1A & 04:39:35.202 & +25:41:44.22 & I & 137 & 183 & 2.3 & 0.56$\pm$0.05 & 30 & 166 & 1, 17\\
B335 & 19:37:00.900 & +07:34:09.81 & 0 & 165 & 41 & 1.4 & 0.15$\pm$0.1 & 8 & 73 & 1, 18, 19\\
\enddata
\tablenotetext{a}{Not observable with JCMT.}
\tablecomments{Coordinates, Class, distances, bolometric temperatures ($T_{\rm bol}$), bolometric luminosities ($L_{\rm bol}$), disk radii ($R_{\rm disk}$), position angles of the disk rotational axes ($\theta_{\rm rot}$), and dynamically determined stellar masses ($M_\star$) are from the eDisk publications and forthcoming papers, except for $M_\star$ of TMC-1A and B335. In the eDisk program, $M_\star$ is estimated by fitting Keplerian rotation to the velocity profile extracted from the position--velocity diagram along the disk major axis observed in the emission lines, while the disk mass is negligible in this analysis, which is found to be smaller than 5\%--10\% of $M_\star$ with the 1.3 mm continuum emission. In the binaries, Ced110~IRS4, Oph~IRS43, R~CrA~IRS7B, and R~CrA~IRAS~32, the Class, $T_{\rm bol}$, and $L_{\rm bol}$ can only be determined for the entire system but not for individual companions, so the same values are listed for both primary and secondary sources.}
\tablerefs{$^1$\citet{Ohashi23}; $^2$\citet{Yamato23}; $^3$Phuong et al.\ (in prep.); $^4$Han et al.\ (in prep.); $^5$\citet{Lin23}; $^6$\citet{van't Hoff23}; $^7$\citet{Sai23}; $^8$Gavino et al.\ (submitted); $^9$\citet{Thieme23}; $^{10}$Santamar{\'i}a-Miranda et al.\ (in prep); $^{11}$\citet{Narayanan23}; $^{12}$\citet{Aso23}; $^{13}$\citet{Flores23}; $^{14}$\citet{Kido23}; $^{15}$\citet{Sharma23}; $^{16}$\citet{Encalada24}; $^{17}$\citet{Aso15}; $^{18}$\citet{Yen10}; $^{19}$\citet{Evans23}}
\end{deluxetable*}

\begin{deluxetable*}{lccccccc}
\tablecaption{JCMT observations and results} \label{tab:jcmt_data}
\tablewidth{0pt}
\tablehead{
\colhead{Source name} & \colhead{$\bar{B_{\theta}}$} & \colhead{$\delta B_{\theta}$} & \colhead{$\theta_{\rm mis}$} & \colhead{$\sigma_\theta$} & \colhead{$I_{\rm peak}$} & \colhead{$I_{\rm rms}$} & \colhead{JCMT}\\
\colhead{} &
\colhead{(\arcdeg)} & \colhead{(\arcdeg)} & \colhead{(\arcdeg)} & \colhead{(\arcdeg)} & \colhead{(Jy Beam$^{-1}$)} & \colhead{(mJy Beam$^{-1}$)} & \colhead{Program ID\tablenotemark{a}}}
\startdata
L1489~IRS & 65$\pm$5 & $<$10 & 89$\pm$4 & 10 & 0.32 & 4 & 1, 2, 3 \\
IRAS~04166+2706\tablenotemark{b} & 45$\pm$3 & 10$\pm$3 & 13$\pm$3 & 9 & 0.45 & 3 & 4 \\
IRAS~04169+2702\tablenotemark{b} & 111$\pm$18 & 13$\pm$4 & 62$\pm$18 & 9 & 0.44 & 3 & 4 \\
IRAS~04302+2247 & 25$\pm$5 & 18$\pm$5 & 60$\pm$5 & 10 & 0.4 & 3 & 2, 5, 6 \\
L1527~IRS\tablenotemark{b} & 28$\pm$7 & 19$\pm$6 & 64$\pm$7 & 9 & 0.93 & 7 & 7, 8, 9 \\
IRAS~15398$-$3359 & 1$\pm$2 & 22$\pm$3 & 26$\pm$2 & 8 & 0.97 & 5 & 3, 6, 10 \\
GSS30~IRS3\tablenotemark{b} & 82$\pm$1 & 8$\pm$2 & 63$\pm$1 & 3 & 0.59 & 6 
& 4 \\
Oph~IRS43~VLA1 & 95$\pm$2 & $<$4 & 49$\pm$2 & 4 & 0.35 & 7 & 2, 5, 11 \\
Oph~IRS43~VLA2 & 95$\pm$2 & $<$4 & 54$\pm$2 & 4 & 0.35 & 7 & 2, 5, 11 \\
IRAS~16253$-$2429\tablenotemark{b} & 138$\pm$16 & \nodata & 65$\pm$16 & 13 & 0.21 & 4 & 12, 13 \\
Oph~IRS63 & 45$\pm$13 & 32$\pm$5 & 14$\pm$13 & 11 & 0.74 & 8 & 5 \\
IRAS~16544$-$1604 & 73$\pm$3 & 10$\pm$3 & 62$\pm$3 & 9 & 0.43 & 4 & 5, 6, 10 \\
R~CrA~IRS5N & 131$\pm$2 & 6$\pm$2 & 40$\pm$2 & 5 & 0.73 & 9 & 2, 5 \\
R~CrA~IRS7B-a & 100$\pm$2 & 14$\pm$1 & 75$\pm$2 & 6 & 2.31 & 9 & 2, 5 \\
R~CrA~IRS7B-b & 100$\pm$2 & 14$\pm$1 & 75$\pm$2 & 6 & 2.31 & 9 & 2, 5 \\
R~CrA~IRAS~32A & 165$\pm$3 & 24$\pm$4 & 60$\pm$3 & 8 & 0.66 & 4 & 6, 11 \\
R~CrA~IRAS~32B & 165$\pm$3 & 24$\pm$4 & 57$\pm$3 & 8 & 0.66 & 4 & 6, 11 \\
TMC-1A & 89$\pm$4 & 14$\pm$4 & 77$\pm$4 & 9 & 0.63 & 4 & 1, 2 \\
B335\tablenotemark{b} & 112$\pm$3 & 26$\pm$9 & 39$\pm$3 & 6 & 1.56 & 5 & 14 \\
\enddata
\tablenotetext{a}{Program ID: $^1$M20BP007; $^2$M21BH20A; $^3$M22AP025; $^4$M16AL004; $^5$M22AH29C; $^6$M22BP056; $^7$M17AP073; $^8$M17BP058; $^9$M21BP074; $^{10}$M21AP016; $^{11}$M21AH20A; $^{12}$M19BP030; $^{13}$M20AP052; $^{14}$M17AP067.}
\tablenotetext{b}{Archival data. Except for IRAS~16253$-$2429, the data has been presented in \citet{Kwon18}, \citet{Yen19,Yen20, Yen21}, and \citet{Eswaraiah21}. 
}
\tablecomments{$\bar{B_{\theta}}$ and $\delta B_{\theta}$ are the mean orientation and angular dispersion of the magnetic fields in the dense cores, and $\theta_{\rm mis}$ is the misalignment angle between the magnetic field and the disk rotational axis. $\sigma_\theta$ is the median uncertainty of the magnetic field orientations in the dense cores. $I_{\rm peak}$ is the 850 $\mu$m peak intensity of the dense core. $I_{\rm rms}$ is the noise level estimated in the 850 $\mu$m Stokes {\it I} maps.}
\end{deluxetable*}

\section{Observations}\label{sec:obs}
Of the 19 eDisk targets, 
we retrieved the POL-2 850\,$\mu$m data of 6 targets from the archive and conducted new observations toward 10 targets between 2020 September 26 and 2023 January 15 (Table~\ref{tab:sample} and \ref{tab:jcmt_data}). 
Three eDisk targets are too far south to be observed by JCMT.  
These POL-2 observations were conducted in the Band 1 and 2 weather conditions (225 GHz opacity $<$ 0.08) with the daisy observing mode at a scanning speed of 8$\arcsec$ s$^{-1}$.
The effective field of view of the POL-2 observations is 3$\arcmin$ where the exposure time exceeds 80\% of the total on-source time, 
and the entire field extends to 16$\arcmin$ in diameter with the exposure time decreasing outwards\footnote{\url{https://www.eaobservatory.org/jcmt/instrumentation/continuum/scuba-2/pol-2/}}.  
The angular resolution is 14$\farcs$6. 
The absolute flux uncertainty is approximately 10\% \citep{Mairs21}.

All our POL-2 data were reduced using the software {\it Starlink} version 2021A \citep{Currie14} and its task {\it pol2map}, following the standard procedures$\footnote{\url{https://www.eaobservatory.org/jcmt/science/reductionanalysis-tutorials/pol-2-dr-tutorial-1/}}$, with the default pixel size of 4$\arcsec$.
The instrumental polarization was corrected in the procedures. The noise levels in the resulting Stokes {\it I} maps range from 3 to 9 mJy Beam$^{-1}$ (Table \ref{tab:jcmt_data}).
Then the Stokes {\it IQU} maps were binned to have a pixel size of 12$\arcsec$, approximately matching the angular resolution, to extract polarization detections.
The polarized intensity ($PI$) was debiased as $PI = \sqrt{Q^2+U^2 - {\sigma_{Q,U}}^2}$, where $\sigma_{Q,U}$ is the weighted mean of the variances on {\it Q} and {\it U}.
The detection criteria for polarized emission were set at a signal-to-noise ratio (S/N) of Stokes {\it I} above 5, S/N of $PI$ above 2, and polarization percentage below 30\%.
Thus, the uncertainty of the polarization orientations ($\sim \frac{1}{2}\frac{\sigma_{Q,U}}{PI}$) for individual detections is smaller than 14$\arcdeg$.
Finally, the magnetic field orientations were inferred by rotating the polarization orientations by 90$\arcdeg$, 
assuming that the shorter axes of dust grains are aligned with the magnetic field. 

\section{Results}\label{sec:results}
Figure~\ref{fig:pol2_map1} and \ref{fig:pol2_map2} present the 850\,$\mu$m continuum maps with a pixel size of 4$\arcsec$ and inferred magnetic field orientations from the JCMT POL-2 data. 
The dense cores of L1489~IRS, GSS30~IRS3, R~CrA~IRS7B and IRS5N, and Oph~IRS43 are embedded in large-scale clumps. 
For these dense cores, we used the dendrogram algorithm and adopted the leaf structures to define their areas with the python package {\it astrodendro}, 
which identifies dense cores by analyzing the hierarchy of the structures \citep{Rosolowsky08}.  
The other dense cores are isolated in the JCMT maps. 
We computed mean Stokes {\it Q} and {\it U} of the polarization detections in the areas of the dense cores. 
For the isolated dense cores, we included all the polarization detections within a radius of 1$\arcmin$ from the protostars to compute mean Stokes {\it Q} and {\it U}.
Then the mean magnetic field orientations in the dense cores were derived from their mean Stokes {\it Q} and {\it U}. 
The derived orientations are insensitive to the exact core areas. 
The uncertainties of the mean orientations were estimated by the error propagation from all the included polarization detections (Table~\ref{tab:jcmt_data}).

To estimate the angular dispersions of the magnetic fields in the dense cores, we first applied unsharp masking \citep[e.g.,][]{Pattle17}. 
We smoothed the observed magnetic field orientations over 3 by 3 pixels ($36\arcsec\times36\arcsec$), 
and subtracted the smoothed field orientations from the original field orientations to remove large-scale magnetic field curvatures.
Then the angular dispersions ($\delta B_{\theta}$) were calculated as the standard deviations ($\delta B_{\rm \theta,ob}$) of the residual angles after the subtraction, and were corrected for the uncertainties in the polarization orientations ($\sigma_{\theta}$; Table~\ref{tab:jcmt_data}) as, 
\begin{equation}
\delta B_{\theta} = \sqrt{{\delta B_{\rm \theta,ob}}^2 - {\sigma_{\theta}}^2}.
\end{equation}
Uncertainties in $\delta B_{\theta}$ were estimated from error propagation. 
We note that changing the smoothing scale for unsharp masking would change the resultant angular dispersions, 
but it does not significantly affect the relative difference among the sample sources.
Thus, the choice of the smoothing scale does not affect our study on the correlations. 
We have tested it, and our conclusions remain the same if we adopt a smoothing scale of 5 by 5 pixels ($60\arcsec\times60\arcsec$) for unsharp masking.
In IRAS16253$-$2429, the limited number of polarization detections prevented us from measuring the angular dispersion. 
For L1489~IRS and Oph~IRS43, we found $\delta B_{\rm \theta,ob} < \sigma_{\theta}$, so we adopted $\sigma_{\theta}$ as the upper limit of the angular dispersions (Table~\ref{tab:jcmt_data}). 

\begin{figure*}
\centering
\includegraphics[width=\textwidth]{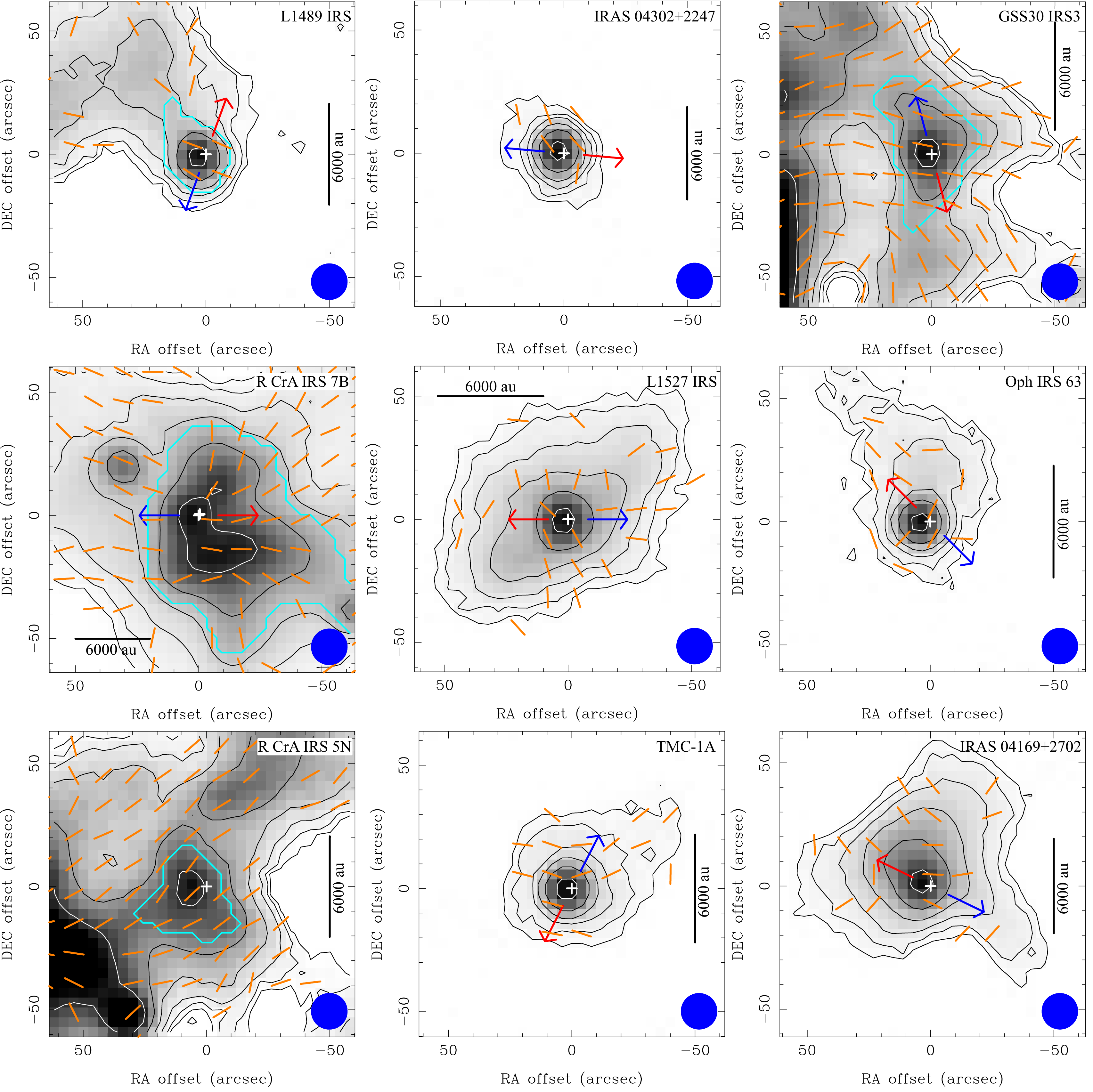}
\caption{JCMT 850\,$\mu$m Stokes {\it I} intensity maps of the eDisk targets. 
Crosses denote the protostellar positions measured with the eDisk program \citep{Ohashi23}. 
Orange segments present the magnetic field orientations inferred from the JCMT POL-2 data with polarization detections above 2$\sigma$, 
and the uncertainties in their orientations are all smaller than 14$\arcdeg$.
Blue filled circles show the angular resolution of 14$\farcs$6.
Red and blue arrows denote the directions of the known blueshifted and redshifted outflows.
Contour levels start from 2.5\% of the peak intensity (Table~\ref{tab:jcmt_data}) and increase in steps of a factor of 2. 
For L1489~IRS, GSS30~IRS3, and R~CrA~IRS7B and IRS5N, 
the cyan contours outline the core areas identified with {\it dendrogram}. 
The maps are arranged from top to bottom and from left to right in descending order of the disk sizes.
}\label{fig:pol2_map1}
\end{figure*}

\begin{figure*}
\centering
\includegraphics[width=\textwidth]{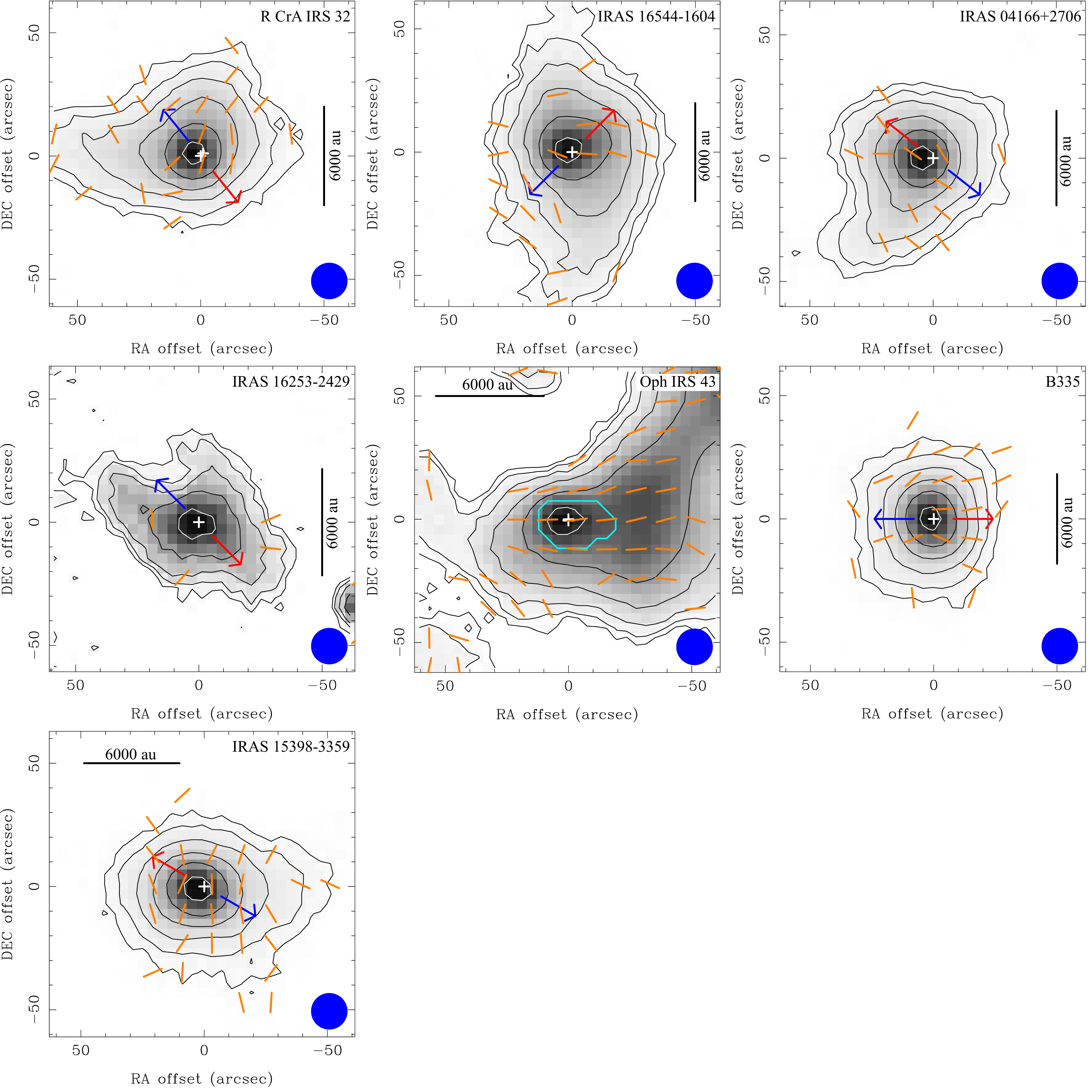}
\caption{Continuation of Fig.~\ref{fig:pol2_map1}. The contour levels in the map of IRAS~16253$-$2429 start from 5\% of the peak intensity. 
For Oph~IRS43, 
the cyan contour outlines the core area identified with {\it dendrogram}. 
}\label{fig:pol2_map2}
\end{figure*}

\section{Analysis}\label{sec:analysis}
We compared the above measurements with the sizes of the protostellar disks in our targets. 
The disk sizes and orientations were estimated from the eDisk 1.3\,mm continuum results, using 2D Gaussian fitting to the continuum emission \citep{Ohashi23}. 
The orientation of the disk minor axis was adopted as the disk rotational axis projected on the plane of the sky, which can represent the projected direction of the net angular momentum of the material that has been accreted onto the star+disk system from the dense core. 
The disk radius was defined as the 2$\sigma$ width of the major axis of the fitted deconvolved 2D Gaussian function, which is equivalent to the radius enclosing 90\% of the total flux with the curve-of-growth method for disks with a Gaussian intensity distribution \citep{Ansdell18}.
The errors from the Gaussian fitting are $<$1--5\%, except for those compact disks around secondary companions, 
so they are negligible compared to the uncertainties of other parameters. 

In Class II disks, the sizes in the millimeter continuum tend to be smaller than those in the CO emission due to dust evolution or differences in optical depths between the continuum and CO emission \citep{Ansdell18}. 
The discrepancy between dust and gas disk radii can be a factor of two or larger. 
In the case of Class 0 and I protostars, where the disk radii have been measured from the radial profiles of rotational velocities (e.g., L1527~IRS and TMC-1A), their disk radii measured in the continuum are 30\%--70\% of those from gas rotation \citep{Aso15, Aso17, van't Hoff23}.
Synthetic images from MHD simulations of disk formation in dense cores suggest that the disk radius determined by gas rotation is typically two to three times larger than the half-width at half-maximum of the fitted Gaussian function to the continuum emission \citep{Aso20}, and thus would be comparable to our defined disk radius, the 2$\sigma$ Gaussian width.
The impact of this possible underestimate of the disk sizes on correlations is discussed in Section~\ref{sec:discussion}.

\begin{figure*}
\centering
\includegraphics[width=\textwidth]{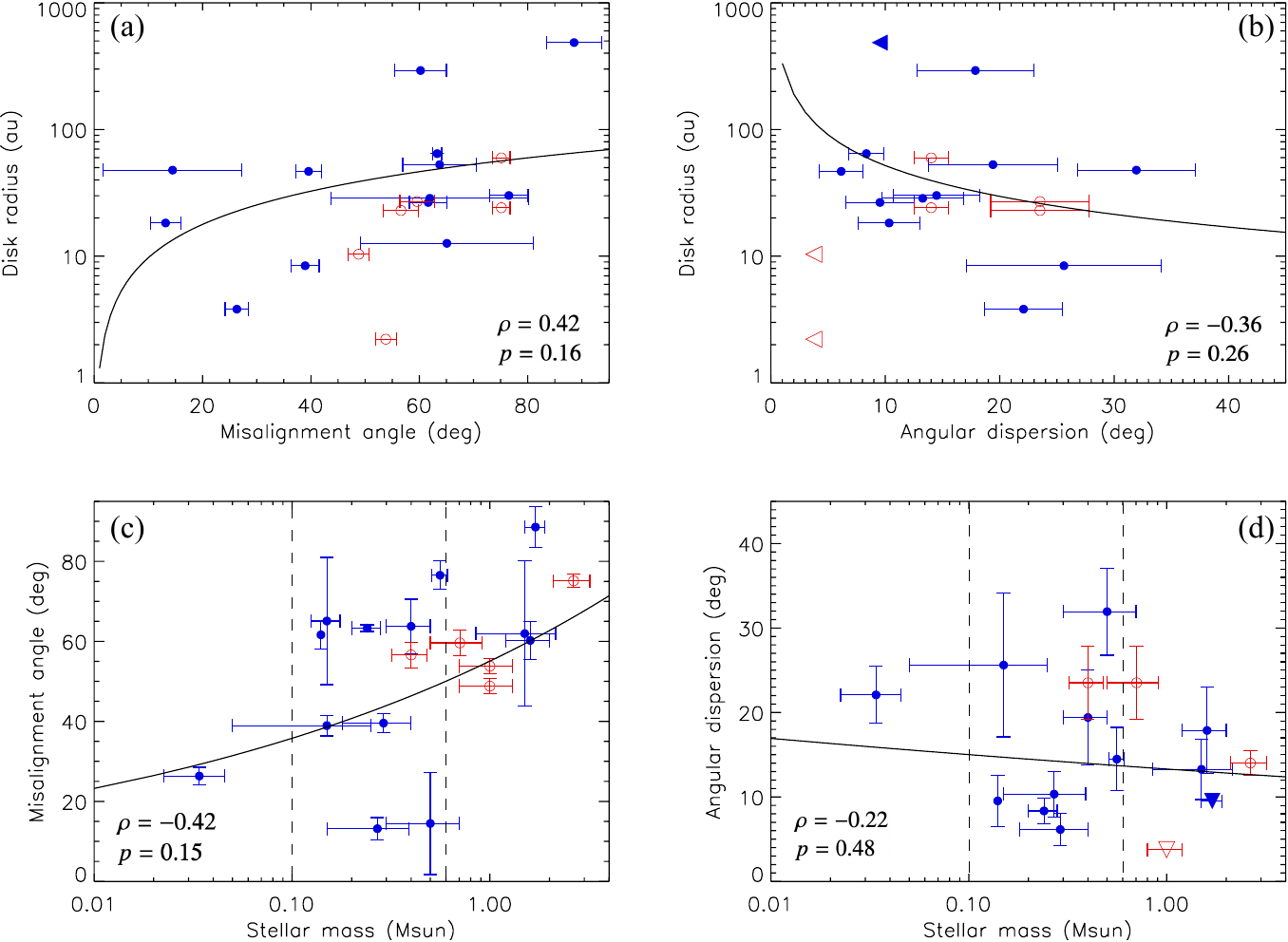}
\caption{Comparisons of the radii of the 1.3\,mm continuum disks with (a) the misalignment angles between the disk rotational axes and the magnetic fields in the dense cores and (b) with the angular dispersions of the magnetic fields. Blue and red data points represent the isolated disks and the disks in the binaries, respectively. Triangles indicate the upper limits of the angular dispersions. Solid curves represent fitted power-law functions to the blue data points. Panel (c) and (d) compare the misalignment angles and angular dispersions with the stellar masses. Dashed lines label the mass range of 0.1--0.6\,$M_\sun$. The values of $\rho$ and $p$ presented in these panels represent the Spearman correlation coefficient and its corresponding $p$-value between the parameter pairs displayed in each panel, respectively.
}\label{fig:Rd_ang}
\end{figure*}

Figure~\ref{fig:Rd_ang}a compares the disk radii with the misalignment angles, which are the angles between the disk rotational axes and the magnetic fields in the dense cores projected on the plane of the sky. 
Blue and red data points represent isolated disks and disks in binaries with separations smaller than 1500\,au, respectively. 
The disks in binaries can be affected by tidal truncation and become smaller \citep{Manara19}, 
so they are included in the plot for comparison but not further discussed regarding a possible correlation.
Figure~\ref{fig:Rd_ang}a shows a tendency that the dense cores with larger misalignment angles harbor larger disks. 
The Spearman's rank correlation coefficient is 0.42 with a {\it p}-value of 0.16, suggesting a tentative correlation between disk radii and misalignment angles, but it is not statistically significant. 
A power-law fit yields an index of 0.9$^{+0.3}_{-0.4}$.

Figure~\ref{fig:Rd_ang}b compares the disk radii and the angular dispersions of the magnetic fields in the dense cores.
The larger disks tend to be associated with the magnetic fields having smaller angular dispersions.
The Spearman correlation coefficient is calculated to be $-0.36$ with a {\it p}-value of 0.26, which is even less statistically significant than the tentative correlation with the misalignment angles.
A power-law fit yields an index of $-0.8^{+0.2}_{-0.3}$.

\begin{figure*}
\centering
\includegraphics[width=\textwidth]{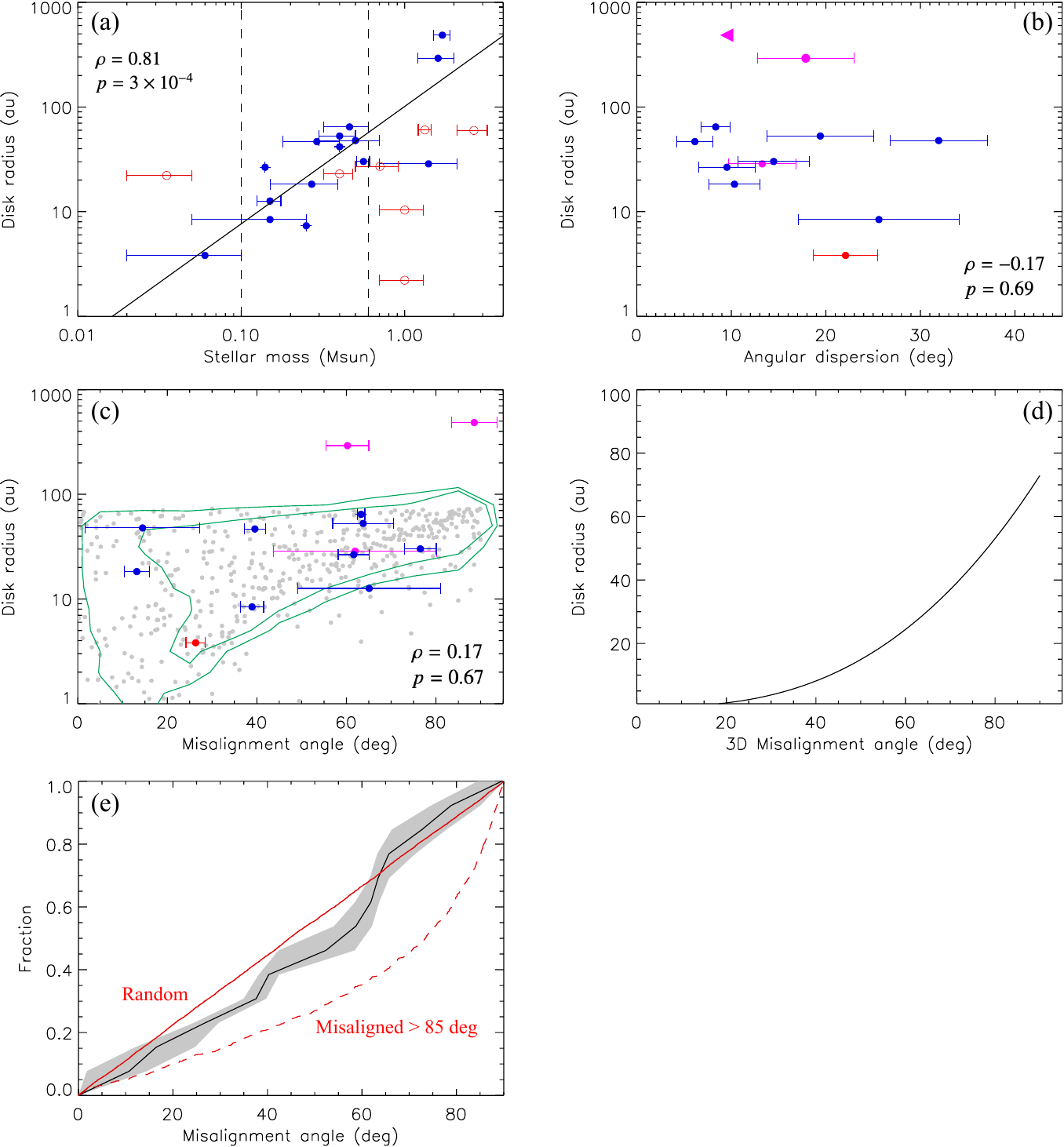}
\caption{Comparison of the radii of the 1.3\,mm continuum disks with (a) the stellar masses, (b) the angular dispersions, and (c) the misalignment angles.
In panel (a), blue and red data points represent the isolated disks and the disks in the binaries, respectively, and the solid curve represents the fitted power-law function to the blue data points. In panel (b) and (c), only the isolated disks are plotted. Blue data points represent the disks with similar stellar masses of 0.1--0.6\,$M_\sun$, while magenta and red data points are the disks with the largest ($>$0.6\,$M_\sun$) and smallest ($<$0.1\,$M_\sun$) stellar masses, respectively. Triangles indicate the upper limits of the angular dispersions. The values of $\rho$ and $p$ presented in panel (a), (b), and (c) represent the Spearman correlation coefficient and its corresponding $p$-value between the parameter pairs displayed in each panel, respectively. Panel (d) shows a correlation between disk radii and 3D misalignment angles that can match the observations. The gray data points in panel (c) are mock data drawn from this correlation, and the green contours show the probability distribution of the mock sample at levels of 68\% and 90\%. (e) Cumulative distribution of the observed misalignment angles in our sample of single protostars (black line) with gray shaded area denoting the associated uncertainty. Red solid and dashed lines illustrate anticipated distributions assuming random orientations of disk rotational axes and magnetic fields, and all misalignment angles exceeding 85$\arcdeg$ in 3D, respectively.
}\label{fig:Rd_Ms}
\end{figure*}

A significant correlation is found between the radii of the isolated disks and the stellar masses having a Spearman correlation coefficient of 0.81 and a {\it p}-value of 0.0003 (Fig.~\ref{fig:Rd_Ms}a). 
The power-law fit gives an index of 1.1$\pm$0.1.
Figure~\ref{fig:Rd_Ms}a additionally shows that the disks in the binaries indeed tend to be smaller than the isolated disks with similar stellar masses.

We also find that in this eDisk sample, the most massive protostars are associated with large misalignment angles and small angular dispersions, and the lowest mass protostars with small misalignment angles and large angular dispersions (Fig.~\ref{fig:Rd_ang}c and d). 
This tendency probably drives the tentative correlations seen in Fig.~\ref{fig:Rd_ang}a and b, 
although the misalignment angles and the angular dispersions are not statistically significantly correlated with the stellar mass, having Spearman correlation coefficients of 0.42 and $-$0.22 and {\it p}-values of 0.15 and 0.48, respectively. 

To exclude any possible dependence of the magnetic field structures on the stellar masses, we compare the disk radii, angular dispersions, and misalignment angles in a subsample with similar stellar masses in the range of 0.1--0.6\,$M_\sun$ in Fig.~\ref{fig:Rd_Ms}b and c (blue data points). 
In this mass range, there is no apparent dependence of  the misalignment angles and angular dispersions on the stellar masses (Fig.~\ref{fig:Rd_ang}c and d).
In Fig.~\ref{fig:Rd_Ms}b and c, magenta and red data points represent the isolated disks with the highest ($>$0.6\,$M_\sun$) and lowest ($<$0.1\,$M_\sun$) stellar masses excluded from this comparison, respectively.
This comparison of the isolated disks with the similar stellar masses shows that their disk radii have no correlation with the misalignment angles and angular dispersions. 
Both have a ${\it p}$ value of 0.7 for Spearman's rank correlation, and the fitted power-law indices are 0.1$^{+0.2}_{-0.1}$ and $-0.3^{+0.2}_{-0.1}$, close to zero.

We note that the misalignment angles in Fig.~\ref{fig:Rd_Ms}c are those projected on the plane of the sky and the projection effect may smooth out a correlation (if any) in Fig.~\ref{fig:Rd_Ms}c.
To examine this possibility, 
we assume a perfect power-law correlation between disk radii and 3D misalignment angles,  
generate a mock sample, and randomly project their 3D misalignment angles on the 2D plane. 
This power-law relationship between disk radii and 3D misalignment angles is an arbitrary choice. 
We vary the power-law index and coefficient to minimize the difference between the observed and mock probability distributions. 
Figure~\ref{fig:Rd_Ms}d illustrates a correlation between disk radii and 3D misalignment angles, resulting in a similar probability distribution to the observations. 
In Fig.~\ref{fig:Rd_Ms}c, the contours represent the probability distribution from this correlation after projection, while the gray data points show the mock data.

\section{Discussion}\label{sec:discussion}
The rarity of large protostellar disks ($>$100 au) and the presence of very small disks ($\lesssim$10 au) around young protostars have often been interpreted as signs of magnetic fields suppressing disk growth via magnetic braking \citep{Yen15, Maury19, Tobin20}.
On the other hand, it is also suggested that magnetic braking needs to be alleviated in collapsing dense cores in order to form sizable disks of tens of au, which are often observed around young protostars, 
and several mechanisms have been proposed \citep{Tsukamoto23}.
Our results reveal a significant correlation between the disk radii and the stellar masses around Class 0 and I protostars in the magnetized dense cores (Fig.~\ref{fig:Rd_Ms}a). 
Despite the disk radii being measured in the 1.3\,mm continuum and potentially underestimating the actual disk sizes by a factor of two, this correlation remains valid. 
We have conducted a power-law fit considering the continuum disk radii as lower limits, and the fitting result remains unchanged. 
Our results provide further constraints on the disk growth process in the presence of magnetic fields, assuming that our sample sources having a variety of stellar masses represent an evolutionary sequence.

The correlation between disk radii and stellar masses is theoretically expected from both HD and non-ideal MHD \citep{Terebey84,Basu98,Hennebelle16,Lee21,Lee24}. 
From HD, there is no magnetic braking, and the angular momentum  is conserved in collapsing dense cores.
This correlation is anticipated to be steep, with a power-law index $\geq$1, because more angular momentum is transported to and accumulated in the disk-forming region with the proceeding collapse, particularly for dense cores having increasing specific angular momentum with increasing radius initially.
From non-ideal MHD, magnetic braking can become inefficient due to imperfect coupling between magnetic fields and matter, and the disk growth is only partially suppressed by magnetic fields. Thus, this correlation is also expected, but it would be shallower, with a power-law index $\leq$0.5, compared to that from HD.
In addition, in both scenarios, HD or non-ideal MHD, disks can be small at the early stage of the gravitational collapse because the majority of the dense core has not collapsed and only limited amount of angular momentum has been transported inward with the collapse.
This picture is similar to our observed correlation showing smaller disks around protostars with lower masses.
Thus, our finding of a steep correlation with a power-law index of 1.1$\pm$0.1 in our targets may suggest inefficient magnetic braking within collapsing dense cores that enables rapid disk growth with the proceeding collapse and formation of sizable disks of tens of au or larger.
Further studies including magnetic field strength and diffusivity are required to fully determine the dependence of disk growth.

Our results also show that the disk radii are tentatively correlated with the projected misalignment angles between the disk rotational axes and the magnetic fields in the dense cores and with the angular dispersions of the magnetic fields (Fig.~\ref{fig:Rd_ang}a and b).
However, when considering a subsample with similar stellar masses to exclude the dependence of the disk radii on the stellar masses, 
no clear dependence is observed (Fig.~\ref{fig:Rd_Ms}b and c). 
These results suggest that the magnetic field orientations and structures on the scale of dense cores have limited influence on the disk sizes. 
The scattering of disk radii in the subsample could be caused by factors such as the angular momentum, magnetic diffusivity, and magnetic field strength in the dense cores \citep{Lee21}. 
Nonetheless, when accounting for the projection effect, the possibility of a positive correlation between the disk radii and misalignment angles in 3D space cannot be ruled out (Fig.\ref{fig:Rd_Ms}c and d), which agrees with theoretical expectations \citep{Hennebelle09,Joos12,Li13,Hirano20}.

In the 13 single protostars in our sample, the misalignment angles range from 10$\arcdeg$ to 90$\arcdeg$, 
and 11 of them have disks larger than 10\,au (Fig.~\ref{fig:Rd_ang}a).
The cumulative distribution of their misalignment angles is consistent with the expectation from random 3D orientations of disk rotational axes and magnetic fields, 
and is distinct from a scenario where all sources have misalignment angles larger than 85$\arcdeg$ (Fig.~\ref{fig:Rd_Ms}e). 
Thus, our results show that sizable protostellar disks can form without largely misaligned rotational axes and magnetic fields.
In ideal MHD simulations with a typical mass-to-flux ratio of 2--3 in dense cores \citep{Crutcher12}, 
sizable Keplerian disks typically form when the misalignment angle is almost 90$\arcdeg$.
If the magnetic field and rotation axis are only slightly misaligned, a weak magnetic field with a mass-to-flux ratio greater than 5--10 would be required to form a sizable Keplerian disk in the ideal MHD limit \citep{Hennebelle09,Joos12,Li13}.
Consequently, even though misalignment between the magnetic field and rotational axis in a dense core may facilitate disk growth, given the potential correlation between 3D misalignment angles and disk sizes in our sample (Fig.~\ref{fig:Rd_Ms}c \& d), our results could suggest that the misalignment is not the primary mechanism driving the formation of sizable Keplerian disks \citep[also see ][]{Kwon18,Yen21a, Yen21}, 
unless the magnetic field is much weaker than generally assumed in theoretical studies. 
Instead, the disk size is more sensitive to the amount of mass that has been accreted onto the star+disk system (Fig.~\ref{fig:Rd_Ms}a), possibly related to magnetic decoupling from the gas \citep{Tu23}.
Estimating the magnetic field strengths and diffusivities in the dense cores is needed to further investigate the origin of this dependence. 

Besides, although the two largest disks in our sample, L1489~IRS and IRAS~04302+2247, are indeed associated with larger misalignment angles ($>$60$\arcdeg$),  
the presence of these large disks may not be solely attributed to their larger misalignment. 
Their stellar masses, which are also highest in our sample, are a factor of 5 higher than the current masses of their dense cores \citep[0.03--0.04\,$M_\sun$;][]{Motte01}.
Therefore, they are at the later evolutionary stage, where the dense cores have been mostly consumed or dissipated. 
In this phase, the magnetic field's impact on protostellar disks is expected to be minimal, and magnetic braking on the disks becomes inefficient \citep{Machida11}. 
As a result, these disks could grow in size more rapidly. 
Furthermore, in these sources, the dynamically collapsing region can extend to a 0.1 pc scale \citep{Shu77,Sai22}.
It is possible that the observed magnetic fields in these dense cores even on a 0.1 pc scale have been dragged around by the gas motions, which could result in the large misalignment angles, and may not represent the initial misalignment. 
Thus, these cases also support that the amount of the accreted mass is a more important factor in the disk growth. 

\section{Summary}
To investigate the disk formation and growth in the presence of magnetic fields, we utilized the JCMT POL-2 data at 850 $\mu$m and studied the magnetic field structures on a 0.1 pc scale in protostellar dense cores harboring the targets in the ALMA eDisk program, where the disk sizes have been measured in the 1.3 mm continuum emission and the protostellar masses from Keplerian disk rotation in the line emission. 
We compared the magnetic field orientations with respect to the disk rotational axes and the magnetic field angular dispersions in the dense cores with the disk sizes and protostellar masses among 15 Class 0 and I single protostars. 

In this sample, we observe a significant correlation between the disk radii and the stellar masses with a power-law index of 1.1$\pm$0.1 within the stellar mass range of 0.05--2 $M_\odot$, and the disks can grow to $>$100 au in radius when the central protostars exceed 1 $M_\odot$. 
We do not find any statistically significant dependence of the disk radii on the projected misalignment angles between the rotational axes of the disks and the magnetic fields in the dense cores, nor on the angular dispersions of the magnetic fields within these cores.
Besides, we find that the observed distribution of the projected misalignment angles in this sample is consistent with the random orientations of the magnetic fields with respect to the disk rotational axes in 3D space. 

These results suggest that the amount of mass that has been accreted onto star+disk systems is a more dominant factor affecting the sizes of protostellar disks, 
while the orientations and structures of magnetic fields in dense cores have a limited impact on the disk formation process. 
In addition, protostellar disks with various sizes form regardless of the orientations of the magnetic fields with respect to the disk rotational axes. 

The observed correlation between the disk radii and the stellar masses could imply that magnetic braking is alleviated within collapsing dense cores, 
and thus disks can grow in size with the proceeding collapse, as expected from non-ideal MHD calculations. 
Nonetheless, when considering the projection effect, we cannot rule out a positive correlation between disk radii and misalignment angles in 3D space.
Thus, a larger misalignment between the magnetic field and rotational axis in a dense core may facilitate disk growth, even though the misalignment is not a primarily mechanism driving the formation of sizable disks.

\begin{acknowledgements}
The James Clerk Maxwell Telescope is operated by the East Asian Observatory on behalf of The National Astronomical Observatory of Japan; Academia Sinica Institute of Astronomy and Astrophysics; the Korea Astronomy and Space Science Institute; the National Astronomical Research Institute of Thailand; Center for Astronomical Mega-Science (as well as the National Key R\&D Program of China with No. 2017YFA0402700). Additional funding support is provided by the Science and Technology Facilities Council of the United Kingdom and participating universities and organizations in the United Kingdom and Canada.
Additional funds for the construction of SCUBA-2 were provided by the Canada Foundation for Innovation.
The authors wish to recognize and acknowledge the very significant cultural role and reverence that the summit of Maunakea has always had within the indigenous Hawaiian community.  We are most fortunate to have the opportunity to conduct observations from this mountain.
This research made use of {\it astrodendro}, a Python package to compute dendrograms of Astronomical data (\url{http://www.dendrograms.org/}).
H.-W.Y.\ acknowledges support from National Science and Technology Council (NSTC) in Taiwan through grant NSTC 110-2628-M-001-003-MY3 and from the Academia Sinica Career Development Award (AS-CDA-111-M03).
JPW acknowledges support from NSF AST-2107841.
PMK acknowledges support from NSTC 108-2112- M-001-012, NSTC 109-2112-M-001-022 and NSTC 110-2112-M-001-057.
J.K.J.\ acknowledge support from the Independent Research Fund Denmark (grant No. 0135-00123B).
W.K.\ was supported by the National Research Foundation of Korea (NRF) grant funded by the Korea government (MSIT) (NRF-2021R1F1A1061794).
C.W.L.\ is supported by the Basic Science Research Program through NRF funded by the Ministry of Education, Science and Technology (NRF- 2019R1A2C1010851), and by the Korea Astronomy and Space Science Institute grant funded by the Korea government (MSIT; Project No. 2023-1-84000). 
ZYL is supported in part by NASA 80NSSC20K0533 and NSF AST-2307199 and AST-1910106.
LWL acknowledges support from NSF AST-2108794.
N.O.\ acknowledges support from NSTC in Taiwan through the grants NSTC 109-2112-M-001-051 and 110-2112-M-001-031.
S.T.\ is supported by JSPS KAKENHI Grant Numbers 21H00048 and 21H04495. This work was supported by NAOJ ALMA Scientific Research Grant Code 2022-20A.
J.J.T.\ acknowledges support from NASA RP 80NSSC22K1159. The National Radio Astronomy Observatory is a facility of the National Science Foundation operated under cooperative agreement by Associated Universities, Inc.
IdG acknowledges support from grant PID2020-114461GB-I00, funded by MCIN/AEI/10.13039/501100011033.
SPL acknowledges grants from NSTC of Taiwan 106-2119-M-007-021-MY3 and 109-2112-M-007-010-MY3.
JEL is supported by NRF grant funded by the Korean government (MSIT) (grant number 2021R1A2C1011718).
\end{acknowledgements}

\software{Starlink \citep{Currie14}, astrodendro \citep{Robitaille19}}

\end{document}